\documentclass[seceq]{ptptex}

\newcommand{\eqdef}{\stackrel{\text{def}}{=}}
\newcommand{\n}{\nonumber \\}
\newcommand{\bm}{\boldsymbol}
\newcommand{\ignore}[1]{}
\newcommand{\paragraph}[1]{\medskip\noindent{\bf #1}\quad}

\title{
Crum's Theorem for `Discrete' Quantum Mechanics
}

\author{
Satoru \textsc{Odake}$^{1}$ and Ryu \textsc{Sasaki}$^{2}$
}

\inst{
$^1$Department of Physics, Shinshu University,
Matsumoto 390-8621, Japan\\
$^2$Yukawa Institute for Theoretical Physics, Kyoto University,\\
Kyoto 606-8502, Japan
}

\abst{
In one-dimensional quantum mechanics, or the Sturm-Liouville theory,
Crum's theorem describes the relationship between the original and
the associated Hamiltonian systems, which are iso-spectral except for
the lowest energy state. Its counterpart in `discrete' quantum
mechanics is formulated algebraically, elucidating the basic structure
of the discrete quantum mechanics, whose Schr\"odinger equation is
a difference equation.
}

\begin{document}

\maketitle

\section{Introduction}
\label{intro}

In the seminal paper of 1955, Crum showed \cite{crum}, if rephrased
in the language of quantum mechanics, the existence of an associated
Hamiltonian system for any given one-dimensional quantum mechanical
Hamiltonian system under mild assumptions.
The method or the technique is quite universal and is known under
many different names; the Darboux transformation \cite{Darboux},
the factorisation method \cite{infhul}
or the supersymmetric quantum mechanics \cite{susyqm}.
Crum himself presented his results in the traditional language of
 Sturm-Liouville systems. 
Since Crum's theorem elucidates the generic structure, 
many exactly and quasi-exactly solvable quantum mechanical examples were
constructed by employing Crum's theorem and its modifications combined with 
shape invariance \cite{genden}.
In particular, Adler' modification of Crum's theorem \cite{adler} is quite
general and useful.
It allows to construct an infinitely many exactly solvable potential from any
exactly solvable one.
 
Recently `discrete' quantum mechanics
was introduced by the present authors \cite{os4,os12,os13}.
It is a generalisation of quantum mechanics, in which the
Schr\"odinger equation is a difference equation instead of
differential in the ordinary quantum mechanics.
In other words, the differential operator in the Hamiltonian is replaced
by finite difference operators, either in the pure imaginary or
the real direction.
This is why they are called discrete quantum mechanics.
Many explicit examples of exactly \cite{os6,os7} and quasi-exactly
solvable \cite{os10,newqes} systems have been constructed.
These exactly solvable systems have salient features;
they are solvable both in the Schr\"odinger and the Heisenberg pictures.
The main part of the exact eigenfunctions are the known hypergeometric
orthogonal polynomials of the Askey scheme \cite{askey,ismail,koeswart}.
On the other hand, the exact Heisenberg operator solutions define
the creation and annihilation operators \cite{os6,os7} which,
together with the Hamiltonian, form the dynamical symmetry algebra
of the exactly solvable systems.
The so-called $q$-oscillator algebra \cite{qoscill} is the most
typical dynamical symmetry algebra realised in this way \cite{os11}.
In discrete quantum mechanics, as will be shown shortly, the counterpart
of Crum's theorem holds and it also elucidates the generic structure of
one dimensional systems.
As in the ordinary quantum mechanics, Crum's theorem and its modifications
are useful for constructing various exactly and quasi-exactly solvable
discrete quantum mechanical examples. In particular, its modification
a l\'a Adler \cite{kos} will also provide an infinite
number of exactly solvable ones based on any exactly solvable one. 
When applied to discrete quantum mechanics with real shifts, the modified
Crum's theorem  generates an infinite or finite series of exactly solvable
birth and death processes based on any known exactly solvable one \cite{birth}.
The insight obtained Crum's theorems and their modification, in the
ordinary and discrete quantum mechanics, is essential for the recent
derivation of infinite numbers of shape invariant systems and new
exceptional orthogonal polynomials \cite{os16,os17}.

In this paper, we present the discrete quantum mechanics version of
 Crum's theorem. Under mild assumptions, we construct algebraically
an associated Hamiltonian system for any given one-dimensional
discrete quantum Hamiltonian system.
That is, if a Hamiltonian together with the full discrete energy spectra
and the corresponding eigenfunctions are given, then the associated
Hamiltonian together with the full discrete energy spectra and
the corresponding eigenfunctions are constructed as in the original
Crum's paper \cite{crum}.
This process goes on indefinitely, since the first associated
Hamiltonian system generates the second associated Hamiltonian
system and so on.
Due to the essential distinction between the differential and
difference equations, several technical assumptions are necessary
for the discrete Crum's theorem.
For example, in one-dimensional ordinary quantum mechanics,
the energy spectra are non-degenerate and the oscillation theorem holds.
That is, the $n$-th excited state wavefunction has $n$ zeros. 
These two properties are not necessarily shared by the generic
discrete QM. 
The hermiticity is almost trivial in ordinary QM, but it can only be
proven after the explicit form of the groundstate wavefunction is
obtained in discrete QM \cite{os10,os13}.
Another distinctive feature is the wavefunction. In ordinary QM,
the wavefunction is a complex valued real function, defined on
the real line, or half line or on a line segment.
In one dimension, the wavefunction can be chosen real.
In discrete QM with pure imaginary shifts, the wavefunction undergoes
shift operations $\psi(x)\to \psi(x\pm i\gamma)$, $\gamma\in\mathbb{R}$. 
Thus we require the {\em analyticity\/} of the wavefunction with
its domain including the real axis or a part of it in which the
dynamical variable $x$ is defined.
In spite of these differences, the generation of the associated
Hamiltonian system goes almost parallel in the ordinary and discrete QM,
since the main part is algebraic.

This paper is organised as follows. In section two, Crum's theorem is
recapitulated in some detail in the language and notation of ordinary
quantum mechanics. This explains the underlying logical structure of
the associated Hamiltonian system, which is also shared by the
discrete quantum mechanics version. The discrete version of Crum's
theorem is stated and proved in section three.
The final section is for a summary and comments.

\section{Ordinary Quantum Mechanics}
\label{ordQM}

Let us start with a generic one-dimensional quantum mechanical system
having discrete semi-infinite energy levels only:
\begin{equation}
  0=\mathcal{E}_0 <\mathcal{E}_1 < \mathcal{E}_2 < \cdots.
  \label{semipositive}
\end{equation}
Here we have chosen the constant part of the Hamiltonian so that
the groundstate energy is zero.
Then the Hamiltonian is {\em positive semi-definite\/} and in one
dimension all the energy levels are non-degenerate. 
It is well known in linear algebra that any positive semi-definite
hermitian matrix can be factorised as a product of a certain matrix,
say $\mathcal{A}$, and its hermitian conjugate $\mathcal{A^\dagger}$.
Similar factorisation applies to the present quantum mechanical
Hamiltonian $\mathcal{H}$, which takes a very simple form:
\begin{align}
  \mathcal{H}&=p^2+U(x)
  =p^2+\Bigl(\frac{d\mathcal{W}(x)}{dx}\Bigr)^2
  +\frac{d^2\mathcal{W}(x)}{dx^2},\qquad p=-i\frac{d}{dx},\\[4pt]
  &=\mathcal{A}^{\dagger}\mathcal{A},\qquad\quad
  \mathcal{A}\eqdef\frac{d}{dx}-\frac{d\mathcal{W}(x)}{dx},\quad
  \mathcal{A}^{\dagger}=-\frac{d}{dx}-\frac{d\mathcal{W}(x)}{dx}.
\end{align}
Here a real function $\mathcal{W}(x)\in\mathbb{C}^\infty$
is called a {\em pre-potential\/}
and it parametrises the groundstate wavefunction $\phi_0(x)$,
which has {\em no node\/} and can be chosen real and positive:
\begin{equation}
  \phi_0(x)=e^{\mathcal{W}(x)}.
  \label{prepotdef}
\end{equation}
It is trivial to verify
\begin{equation}
  \mathcal{A}\phi_0(x)=0\ \Rightarrow\ \mathcal{H}\phi_0(x)=0.
\end{equation}
All the eigenfunctions are square-integrable and orthogonal with
each other and form a complete basis of the Hilbert space:
\begin{alignat}{2}
  \mathcal{H}\phi_n(x)&=\mathcal{E}_n\phi_n(x),
  &\quad n&=0,1,2,\ldots,\\
  \int\phi_n(x)^*\phi_m(x)dx&=h_n\delta_{nm},\quad
  0<h_n<\infty,&\quad n,m&=0,1,2,\ldots.
\end{alignat}
It is well-known that the $n$-th excited wavefunction $\phi_n(x)$
has $n$ zeros in the interior.
For simplicity we choose all the eigenfunctions to be real.
Here are a few examples:
\ignore{
\begin{alignat}{3}
  \mathcal{W}(x)&=-\tfrac12x^2,&U(x)&=x^2-1,&
  &\hspace*{-7mm}-\infty<x<\infty,\\
  \mathcal{W}(x)&=-\tfrac12x^2+g\log x,&\quad
  U(x)&=x^2+\frac{g(g-1)}{x^2}-1-2g,\quad&g>1,\quad&0<x<\infty,\\
  \mathcal{W}(x)&=g\log\sin x,&
  U(x)&=\frac{g(g-1)}{\sin^2x}-g^2,\quad&g>1,\quad&0<x<\pi.
\end{alignat}
}
\begin{alignat}{2}
  \mathcal{W}(x)&=-\tfrac12x^2,&U(x)&=x^2-1,
  \qquad\qquad -\infty<x<\infty,\\
  \mathcal{W}(x)&=-\tfrac12x^2+g\log x,&
  \ \ U(x)&=x^2+\frac{g(g-1)}{x^2}-1-2g,\ g>1,\ 0<x<\infty,\\
  \mathcal{W}(x)&=g\log\sin x,&
  U(x)&=\frac{g(g-1)}{\sin^2x}-g^2,\quad g>1,\ \ 0<x<\pi.
\end{alignat}
They all lead to well-known exactly solvable quantum mechanics
whose eigenfunctions consist of the classical orthogonal polynomials,
the Hermite, Laguerre and Jacobi polynomials, respectively.
In each case, the squared groundstate wavefunction $\phi_0(x)^2$
gives the orthogonality weight function for the polynomials.

\bigskip
Next let us define an associated Hamiltonian $\mathcal{H}^{[1]}$ by
simply changing the order of $\mathcal{A}$ and $\mathcal{A}^\dagger$:
\begin{equation}
  \mathcal{H}^{[1]}\eqdef\mathcal{A}\mathcal{A}^\dagger.
  \label{H1def}
\end{equation}
For later convenience, let us attach the superscript ${}^{[0]}$ to
all the quantities in the original Hamiltonian system,
$\mathcal{H}^{[0]}\eqdef\mathcal{H}$,
$\phi^{[0]}_n(x)\eqdef\phi_n(x)$,
$\mathcal{A}^{[0]}\eqdef\mathcal{A}$,
$U^{[0]}(x)\eqdef U(x)$,
$\mathcal{W}^{[0]}(x)\eqdef\mathcal{W}(x)$.
By construction $\mathcal{A}$ and $\mathcal{A}^\dagger$ intertwine
$\mathcal{H}^{[0]}$ and $\mathcal{H}^{[1]}$:
\begin{equation}
  \mathcal{A}^{[0]}\mathcal{H}^{[0]}=\mathcal{H}^{[1]}\mathcal{A}^{[0]},
  \qquad
  \mathcal{A}^{[0]\dagger}\mathcal{H}^{[1]}
  =\mathcal{H}^{[0]}\mathcal{A}^{[0]\dagger}.
  \label{intertwine}
\end{equation}
We will show that the associated Hamiltonian system
$\mathcal{H}^{[1]}$ is {\em iso-spectral\/} to the original Hamiltonian
system $\mathcal{H}^{[0]}$ and the eigenfunctions are in one to one
correspondence, {\em except for\/} the groundstate. 
Thanks to the first of the above relation \eqref{intertwine},
it is trivial to verify that the eigenfunctions of the associated
Hamiltonian system $\mathcal{H}^{[1]}$  are generated algebraically
by multiplying $\mathcal{A}^{[0]}$ to the eigenfunction of the
original system:
\begin{alignat}{2}
  \phi^{[1]}_n(x)\eqdef&\mathcal{A}^{[0]}\phi^{[0]}_n(x),\quad
  \int\phi^{[1]}_n(x)^*\phi^{[1]}_m(x)dx=\mathcal{E}_nh_n\delta_{nm},
  &\quad\ n,m&=1,2,\ldots,\\
  &\mathcal{H}^{[1]}\phi^{[1]}_n(x)=\mathcal{E}_n\phi^{[1]}_n(x),
  &n&=1,2,\ldots.
  \label{H1spec}
\end{alignat}
Suppose the associated Hamiltonian $\mathcal{H}^{[1]}$ has an
eigenfunction $\phi'(x)$ with the eigenvalue $\mathcal{E}'$
other than those listed above:
\begin{equation}
  \mathcal{H}^{[1]}\phi'(x)=\mathcal{E}'\phi'(x).
\end{equation}
Again, thanks to the second of the relation \eqref{intertwine},
it is  trivial to verify
\begin{equation}
  \mathcal{H}^{[0]}\mathcal{A}^{[0]\dagger}\phi'(x)
  =\mathcal{E}'\mathcal{A}^{[0]\dagger}\phi'(x).
\end{equation}
Due to the completeness of the spectrum of the original Hamiltonian
$\mathcal{H}^{[0]}$, the provisional eigenvalue $\mathcal{E}'$ must
belong to the above spectrum \eqref{H1spec} for $n=1,2.\ldots$.
In other words, $\mathcal{E}'$ cannot be vanishing, $\mathcal{E}'\neq0$.
Suppose that is the case ($\mathcal{E}'=0$), then $\phi'$ is annihilated by 
$\mathcal{A}^{[0]\dagger}$.
Surely there exists a solution of a first order differential equation
$\mathcal{A}^{[0]\dagger}\phi'(x)=0$,
$\phi'(x)=1/\phi^{[0]}_0(x)=e^{-\mathcal{W}(x)}$.
But it is obviously non square-integrable and it does not belong to the
Hilbert space of the associated Hamiltonian system.
After Crum\cite{crum}, we can show in the following way that
$\phi^{[1]}_n(x)$ has exactly $n-1$ zeros. 
Since $\phi^{[0]}_n$ has exactly $n$ zeros, the relation
$$
  \frac{\phi^{[1]}_n(x)}{\phi^{[0]}_0(x)}
  =\frac{d}{dx}\Bigl(\frac{\phi^{[0]}_n(x)}{\phi^{[0]}_0(x)}\Bigr)
$$
tells through Rolle's theorem that $\phi^{[1]}_n$ has at least $n-1$ zeros.
From the relation
$$
  \frac{d}{dx}\Bigl(\phi^{[0]}_0(x)\phi^{[1]}_n(x)\Bigr)
  =-\mathcal{E}_n\phi^{[0]}_0(x)\phi^{[0]}_n(x),
$$
we find that $\phi^{[1]}_n$ has at most $n-1$ zeros.
Thus we have established that the associated Hamiltonian system
$\mathcal{H}^{[1]}$ is {\em iso-spectral\/} to the original Hamiltonian
system $\mathcal{H}^{[0]}$ and the eigenfunctions are in one to one
correspondence, {\em except for\/} the groundstate with the wavefunction
$\phi_0^{[0]}(x)$.
If the groundstate energy $\mathcal{E}_1$ is subtracted from the
associated Hamiltonian $\mathcal{H}^{[1]}$, it is again positive
semi-definite and can be factorised as above:
\begin{align}
  &\mathcal{H}^{[1]}=\mathcal{A}^{[1]\dagger}\mathcal{A}^{[1]}+\mathcal{E}_1
  =p^2+U^{[1]}(x)+\mathcal{E}_1,
  \label{H1_rewrite}\\
  &\mathcal{A}^{[1]}\eqdef\frac{d}{dx}-\frac{d\mathcal{W}^{[1]}(x)}{dx},\quad
  \mathcal{A}^{[1]\dagger}=-\frac{d}{dx}-\frac{d\mathcal{W}^{[1]}(x)}{dx},\\
  &e^{\mathcal{W}^{[1]}(x)}\eqdef\bigl|\phi_1^{[1]}(x)\bigr|,\quad
  U^{[1]}(x)\eqdef\Bigl(\frac{d\mathcal{W}^{[1]}(x)}{dx}\Bigr)^2
  +\frac{d^2\mathcal{W}^{[1]}(x)}{dx^2},
  \label{W1def}\\
  &\mathcal{A}^{[1]}\phi^{[1]}_1(x)=0.
\end{align}
As shown above, the groundstate wavefunction $\phi_1^{[1]}(x)$ has no node.
Note that \eqref{H1def}, \eqref{H1_rewrite} and \eqref{W1def} imply
the Riccati equation for $d\mathcal{W}^{[1]}(x)/dx$:
\begin{equation}
  \Bigl(\frac{d\mathcal{W}^{[1]}(x)}{dx}\Bigr)^2
  +\frac{d^2\mathcal{W}^{[1]}(x)}{dx^2}
  =\Bigl(\frac{d\mathcal{W}(x)}{dx}\Bigr)^2
  -\frac{d^2\mathcal{W}(x)}{dx^2}-\mathcal{E}_1.
  \label{riccati}
\end{equation}
Then by reversing the order of $\mathcal{A}^{[1]\dagger}$ and
$\mathcal{A}^{[1]}$ the second associated Hamiltonian system
$\mathcal{H}^{[2]}$ can be defined. This process can go indefinitely.

Here we list the definition of the $s$-th quantities step by step
for $s\geq 1$,
\begin{align}
  &\mathcal{H}^{[s]}\eqdef\mathcal{A}^{[s-1]}\mathcal{A}^{[s-1]\,\dagger}
  +\mathcal{E}_{s-1},\\
  &\phi^{[s]}_n(x)\eqdef\mathcal{A}^{[s-1]}\phi^{[s-1]}_n(x),\quad(n\geq s),\\
  &e^{\mathcal{W}^{[s]}(x)}\eqdef\bigl|\phi^{[s]}_s(x)\bigr|,\\
  &\mathcal{A}^{[s]}\eqdef\frac{d}{dx}
  -\frac{d\mathcal{W}^{[s]}(x)}{dx},\quad
  \mathcal{A}^{[s]\,\dagger}=-\frac{d}{dx}
  -\frac{d\mathcal{W}^{[s]}(x)}{dx},\\
  &U^{[s]}(x)\eqdef\Bigl(\frac{d\mathcal{W}^{[s]}(x)}{dx}\Bigr)^2
  +\frac{d^2\mathcal{W}^{[s]}(x)}{dx^2}.
\end{align}
Then we can show the following for $n\geq s\geq 0$,
\begin{align}
  &\mathcal{H}^{[s]}\phi^{[s]}_n(x)=\mathcal{E}_n\phi^{[s]}_n(x),\\
  &\phi^{[s]}_n(x)\text{ : real function},\\
  &\mathcal{A}^{[s]}\phi^{[s]}_s(x)=0,\\
  &\mathcal{H}^{[s]}=\mathcal{A}^{[s]\,\dagger}\mathcal{A}^{[s]}
  +\mathcal{E}_{s}
  =p^2+U^{[s]}(x)+\mathcal{E}_{s}.
\end{align}
We have also
\begin{equation}
  \phi^{[s-1]}_n(x)=\frac{\mathcal{A}^{[s-1]\,\dagger}}
  {\mathcal{E}_n-\mathcal{E}_{s-1}}\,\phi^{[s]}_n(x),\quad(n\geq s\geq 1).
\end{equation}
In terms of the determinant (Wronskian)
\begin{equation}
  \text{W}\,[f_1,\ldots,f_n](x)
  \eqdef\det\Bigl(\frac{d^{j-1}f_k(x)}{dx^{j-1}}\Bigr)_{1\leq j,k\leq n},
\end{equation}
(for $n=0$, we set $\text{W}\,[\cdot](x)=1$.), we have
\begin{equation}
  \text{W}\,[\phi_0,\phi_1,\ldots,\phi_{s-1},\phi_n](x)
  =\phi_0(x)\phi^{[1]}_1(x)\cdots\phi^{[s-1]}_{s-1}(x)\phi^{[s]}_n(x),
  \quad(n\geq s\geq 0).
  \label{crumwrons1}
\end{equation}
Therefore we arrive at the concise formulas due to Crum \cite{crum}:
\begin{align}
  &\text{W}\,[\phi_0,\phi_1,\ldots,\phi_{s-1}](x)=
  \phi_0(x)\phi^{[1]}_1(x)\cdots\phi^{[s-1]}_{s-1}(x)=\pm
  e^{\mathcal{W}(x)+\mathcal{W}^{[1]}(x)+\cdots+\mathcal{W}^{[s-1]}(x)},\\
  &U^{[s]}(x)=U(x)-2\frac{d^2}{dx^2}\Bigl(
  \log\text{W}\,[\phi_0,\phi_1,\ldots,\phi_{s-1}](x)\Bigr),\\
  &\phi^{[s]}_n(x)=
  \frac{\text{W}\,[\phi_0,\phi_1,\ldots,\phi_{s-1},\phi_n](x)}
  {\text{W}\,[\phi_0,\phi_1,\ldots,\phi_{s-1}](x)},
  \quad(n\geq s\geq 0).
  \label{crumwronsfin}
\end{align}

\section{`Discrete' Quantum Mechanics (pure imaginary shifts)}
\label{discrQM}

In discrete quantum mechanics, the dynamical variables are,
as in ordinary QM, the coordinate $x$, which takes value in an infinite
or a semi-infinite or a finite range of the real axis and the canonical
momentum $p$, which is realised as a differential operator $p=-i\partial_x$.
Since the momentum operator appears in exponentiated forms
$e^{\pm \gamma p}$, $\gamma\in\mathbb{R}$, in a Hamiltonian,
it causes finite pure imaginary shifts in the wavefunction
$e^{\pm \gamma p}\psi(x)=\psi(x\mp i\gamma)$.
This requires the wavefunction as well as other functions appearing
in the Hamiltonian to be {\em analytic\/} in $x$ within a certain domain
including the physical region of the coordinate.
Let us introduce the $*$-operation on an analytic function,
$*:f\mapsto f^*$.
If $f(x)=\sum\limits_{n}a_nx^n$, $a_n\in\mathbb{C}$, then
$f^*(x)\eqdef\sum\limits_{n}a_n^*x^n$, in which $a_n^*$ is the complex
conjugation of $a_n$. Obviously $f^{**}(x)=f(x)$ and $f(x)^*=f^*(x^*)$.
If a function satisfies $f^*=f$, then it takes real values on the real line.

The starting point is again a generic one dimensional discrete
quantum mechanics Hamiltonian with discrete semi-infinite energy levels
only \eqref{semipositive}. 
Again we assume that the groundstate energy is chosen to be zero
$\mathcal{E}_0=0$, so that the Hamiltonian is positive semi-definite.
The generic factorised Hamiltonian reads
\begin{align}
  &\mathcal{H}=\mathcal{A}^{\dagger}\mathcal{A}
  =\sqrt{V(x)}\,e^{\gamma p}\sqrt{V^*(x)}
  +\!\sqrt{V^*(x)}\,e^{-\gamma p}\sqrt{V(x)}-V(x)-V^*(x),
  \label{discrham}\\
  &\mathcal{A}\eqdef i\bigl(e^{\frac{\gamma}{2}p}\sqrt{V^*(x)}
  -e^{-\frac{\gamma}{2}p}\sqrt{V(x)}\,\bigr),
  \ \mathcal{A}^{\dagger}\eqdef -i\bigl(\sqrt{V(x)}\,e^{\frac{\gamma}{2}p}
  -\sqrt{V^*(x)}\,e^{-\frac{\gamma}{2}p}\bigr).
\end{align}
By specifying the function $V(x)$, various explicit examples are
obtained \cite{os4,os13,os7}.
For instance, 
\begin{align}
  V(x)&=\frac{\prod_{j=1}^4(1-a_je^{ix})}{(1-e^{2ix})(1-q\,e^{2ix})},
  \quad 0< x <\pi,\quad 0<q<1,\n
  &\qquad\quad |a_j|<1,\quad
  \{a_1^*,a_2^*,a_3^*,a_4^*\}=\{a_1,a_2,a_3,a_4\}\ (\text{as a set}),
  \label{AskeyWil}
\end{align}
gives an exactly solvable dynamics whose eigenfunctions consist of
the Askey-Wilson polynomials \cite{askey,ismail,koeswart} times
the groundstate wavefunction $\phi_0(x)$ \eqref{discphi0def},
which gives the orthogonality measure $\phi_0(x)^2$.

Let us emphasise that the corresponding Schr\"odinger equation
$\mathcal{H}\psi(x)=\mathcal{E}\psi(x)$ is a difference equation
\begin{align}
  \sqrt{V(x)V^*(x-i\gamma)}\,\psi(x-i\gamma)
  &+\sqrt{V^*(x)V(x+i\gamma)}\,\psi(x+i\gamma)\n
  &-\bigl(V(x)+V^*(x)\bigr)\psi(x)=\mathcal{E}\psi(x),
  \label{maineq}
\end{align}
instead of differential in ordinary QM.
Again the groundstate wavefunction $\phi_0(x)$ is 
determined as a zero mode of $\mathcal{A}$:
\begin{equation}
  \mathcal{A}\phi_0(x)=0\ \Rightarrow\ \mathcal{H}\phi_0(x)=0.
  \label{discphi0def}
\end{equation}
The above equation for $\phi_0$ reads
\begin{equation}
  \sqrt{V^*(x-i\tfrac{\gamma}{2})}\phi_0(x-i\tfrac{\gamma}{2})-
  \sqrt{V(x+i\tfrac{\gamma}{2})}\phi_0(x+i\tfrac{\gamma}{2})=0.
  \label{explizero}
\end{equation}
This dictates how the `phase' of the potential function $V$ is related
to that of the groundstate wavefunction $\phi_0$.
Here we also assume that the groundstate wavefunction $\phi_0(x)$
has no node and chosen to be real and positive for real $x$.

Due to the lack of generic theorems in the theory of difference equations,
let us assume that all the energy levels are non-degenerate and all
the eigenfunctions are square-integrable and orthogonal with each other
and form a complete basis of the Hilbert space:
\begin{alignat}{2}
  \mathcal{H}\phi_n(x)&=\mathcal{E}_n\phi_n(x),
  &\quad n&=0,1,2,\ldots,\\
  \int\phi_n(x)^*\phi_m(x)dx&=h_n\delta_{nm},\quad
  0<h_n<\infty,&\quad n,m&=0,1,2,\ldots.
\end{alignat}
In most explicit examples these statements can be verified straightforwardly.
For simplicity we choose all the eigenfunctions to be real on the real
axis $\phi^*_n=\phi_n$.
This is possible since $\mathcal{H}$ maps a `real' function to a `real'
function $f^*=f\Rightarrow(\mathcal{H}f)^*=\mathcal{H}f$.

Now the procedure to generate the associated Hamiltonian system goes
almost parallel with the one shown in the preceding section.
We define an associated Hamiltonian $\mathcal{H}^{[1]}$ by simply
changing the order of $\mathcal{A}$ and $\mathcal{A}^\dagger$,
$\mathcal{H}^{[1]}\eqdef \mathcal{A}\mathcal{A}^\dagger$,
as in \eqref{H1def}. Then as in \eqref{intertwine}, we have
$\mathcal{A}^{[0]}\mathcal{H}^{[0]}=\mathcal{H}^{[1]}\mathcal{A}^{[0]}$
and $\mathcal{A}^{[0]\dagger}\mathcal{H}^{[1]}
=\mathcal{H}^{[0]}\mathcal{A}^{[0]\dagger}$.
Here again we have indexed the quantities of the original Hamiltonian
system by the superscript ${}^{[0]}$, as before.
The eigenfunctions and the eigenvalues of the associated Hamiltonian
$\mathcal{H}^{[1]}$ are given by
\begin{alignat}{2}
  \phi^{[1]}_n(x)\eqdef&\mathcal{A}^{[0]}\phi^{[0]}_n(x),\quad
  \int\phi^{[1]}_n(x)^*\phi^{[1]}_m(x)dx=\mathcal{E}_nh_n\delta_{nm},
  &\quad\ n,m&=1,2,\ldots,\\
  &\mathcal{H}^{[1]}\phi^{[1]}_n(x)=\mathcal{E}_n\phi^{[1]}_n(x),
  &n&=1,2,\ldots.
  \label{H1spec2}
\end{alignat}
The same argument as before establishes the iso-spectrality and
the one to one correspondence of the eigenfunctions of
$\mathcal{H}^{[0]}$ and $\mathcal{H}^{[1]}$ except for the groundstate.
In explicit examples given in \cite{os13}, one can show that
$\phi'(x)=\phi'(x\,;\bm{\lambda})
=\varphi(x)/\phi_0(x\,;\bm{\lambda}+\bm{\delta})$
is annihilated by $\mathcal{A}^\dagger$, $\mathcal{A}^\dagger\phi'(x)=0$
as in the Darboux transformation.
Such functions, as before, do not belong to the Hilbert space of the
associated Hamiltonian $\mathcal{H}^{[1]}$.
Here $\bm{\lambda}$ and $\bm{\delta}$ are parameters and their shift.
For example, $\varphi\equiv1$ for the Meixner-Pollaczek case,
$\varphi(x)=x$ for the Wilson case and $\varphi(x)=\sin x$ for the
Askey-Wilson case \eqref{AskeyWil}.
We have used the convention to specify the discrete quantum mechanics
by the name of the polynomials constituting the main part of the
eigenfunctions.

Thus $\phi^{[1]}_1(x)$ is the groundstate wavefunction of $\mathcal{H}^{[1]}$
with the eigenvalue $\mathcal{E}_1$.
By subtracting it from $\mathcal{H}^{[1]}$, we can again factorise
the positive semi-definite Hamiltonian:
\begin{align}
  &\mathcal{H}^{[1]}=\mathcal{A}^{[1]\dagger}\mathcal{A}^{[1]}+\mathcal{E}_1,\\
  &\mathcal{A}^{[1]}\eqdef i\bigl(e^{\frac{\gamma}{2}p}\sqrt{V^{[1]*}(x)}
  -e^{-\frac{\gamma}{2}p}\sqrt{V^{[1]}(x)}\,\bigr),\n
  &\mathcal{A}^{[1]\dagger}\eqdef
  -i\bigl(\sqrt{V^{[1]}(x)}\,e^{\frac{\gamma}{2}p}
  -\sqrt{V^{[1]*}(x)}\,e^{-\frac{\gamma}{2}p}\bigr),
\end{align}
with the new potential function $V^{[1]}$ to be determined now.
This imposes quadratic relations between the function $V^{[0]}$ and
the unknown $V^{[1]}$:
\begin{align}
  &V^{[0]}(x-i\tfrac{\gamma}{2})
  V^{[0]*}(x-i\tfrac{\gamma}{2})
  =V^{[1]}(x)V^{[1]*}(x-i\gamma),
  \label{quadratic}\\
  &V^{[0]}(x+i\tfrac{\gamma}{2})
  +V^{[0]*}(x-i\tfrac{\gamma}{2})
  =V^{[1]}(x)+V^{[1]*}(x)-\mathcal{E}_1,
  \label{linear}
\end{align}
which are discrete counterparts of the Riccati equation \eqref{riccati}
for the pre-potential $\mathcal{W}$.
One essential problem is that the connection between the groundstate
wavefunction $\phi_0(x)$ and the function $V(x)$ in the Hamiltonian is
indirect, in contrast to the one $\phi_0(x)=e^{\mathcal{W}(x)}$,
\eqref{prepotdef} in ordinary QM.
In discrete QM, the potential function $V$ is complex (analytic). 
The information on the `absolute value' of $V^{[1]}(x)$ can be extracted
from \eqref{quadratic}.
The `phase' part of of $V^{[1]}(x)$ is given by that of $\phi^{[1]}_1(x)$
through the zero mode equation of $\mathcal{A}^{[1]}$ \eqref{a1annihilate},
just as the `phase' of $\phi_0$ is
related to that of $V^{[0]}$ through the the zero mode
equation of $\mathcal{A}^{[0]}$, \eqref{explizero}.
Thus the following formula determining the function $V^{[1]}(x)$ in terms
of the previous function $V^{[0]}(x)$ and $\phi^{[1]}_1(x)$
is the main result of the present paper:
\begin{align}
  &V^{[1]}(x)\eqdef\sqrt{V^{[0]}(x-i\tfrac{\gamma}{2})
  V^{[0]*}(x-i\tfrac{\gamma}{2})}
  \ \frac{\phi^{[1]}_1(x-i\gamma)}{\phi^{[1]}_1(x)},
  \label{V1}\\
  &\mathcal{A}^{[1]}{\phi^{[1]}_1(x)}=0.
  \label{a1annihilate}
\end{align}
There is no ambiguity in the phase of the square root in the above
expression \eqref{V1}. It is positive for
$x=\alpha+i\tfrac{\gamma}{2}$, $\alpha\in\mathbb{R}$,
for which the function inside the square root sign is positive definite.
Here we have to assume that the groundstate wavefunction $\phi^{[1]}_1(x)$
has no node, so that the function $V^{[1]}(x)$ does not develop unwanted
singularities in the physical region.
It is rather straightforward to verify that \eqref{quadratic}
is actually satisfied by the $V^{[1]}$ in \eqref{V1}.
In order to verify the linear relation \eqref{linear}, one has to use
the eigenvalue equation for $\phi^{[0]}_1$.
By reversing the order of $\mathcal{A}^{[1]\dagger}$ and $\mathcal{A}^{[1]}$,
the second associated Hamiltonian system $ \mathcal{H}^{[2]}$ can be defined.
Again this process can go indefinitely.

Here we list the definition of the $s$-th quantities step by step
for $s\geq 1$,
\begin{align}
  &\mathcal{H}^{[s]}\eqdef\mathcal{A}^{[s-1]}\mathcal{A}^{[s-1]\,\dagger}
  +\mathcal{E}_{s-1},\\
  &\phi^{[s]}_n(x)\eqdef\mathcal{A}^{[s-1]}\phi^{[s-1]}_n(x),\quad(n\geq s),\\
  &V^{[s]}(x)\eqdef\sqrt{V^{[s-1]}(x-i\tfrac{\gamma}{2})
  V^{[s-1]*}(x-i\tfrac{\gamma}{2})}
  \ \frac{\phi^{[s]}_s(x-i\gamma)}{\phi^{[s]}_s(x)},
  \label{Vsformsq}\\
  &\mathcal{A}^{[s]}\eqdef i\bigl(e^{\frac{\gamma}{2}p}\sqrt{V^{[s]*}(x)}
  -e^{-\frac{\gamma}{2}p}\sqrt{V^{[s]}(x)}\bigr),\n
  &\mathcal{A}^{[s]\,\dagger}
  \eqdef -i\bigl(\sqrt{V^{[s]}(x)}\,e^{\frac{\gamma}{2}p}
  -\sqrt{V^{[s]*}(x)}\,e^{-\frac{\gamma}{2}p}\bigr).
\end{align}
Let us note that the phase of the square root in \eqref{Vsformsq} has
no ambiguity.
Then we can show the following for $n\geq s\geq 0$,
\begin{align}
  &\mathcal{H}^{[s]}\phi^{[s]}_n(x)=\mathcal{E}_n\phi^{[s]}_n(x),\\
  &\phi^{[s]*}_n=\phi^{[s]}_n\ \ \text{ : `real' function},\\
  &\mathcal{A}^{[s]}\phi^{[s]}_s(x)=0,\\
  &\mathcal{H}^{[s]}=\mathcal{A}^{[s]\,\dagger}\mathcal{A}^{[s]}
  +\mathcal{E}_{s}.
\end{align}
We have also
\begin{equation}
  \phi^{[s-1]}_n(x)=\frac{\mathcal{A}^{[s-1]\,\dagger}}
  {\mathcal{E}_n-\mathcal{E}_{s-1}}\,\phi^{[s]}_n(x),\quad(n\geq s\geq 1).
\end{equation}

The discrete counterpart of the determinant formulas of Crum
\eqref{crumwrons1}--\eqref{crumwronsfin} requires a deformation of
the Wronskian, the Casorati determinant, which has a good limiting property:
\begin{align}
  &\text{W}_{\gamma}[f_1,\ldots,f_n](x)
  \eqdef i^{\frac12n(n-1)}
  \det\Bigl(f_k(x+i\tfrac{n+1-2j}{2}\gamma)\Bigr)_{1\leq j,k\leq n},\\
  &\lim_{\gamma\to 0}\gamma^{-\frac12n(n-1)}
  \text{W}_{\gamma}[f_1,f_2,\ldots,f_n](x)
  =\text{W}\,[f_1,f_2,\ldots,f_n](x),
\end{align}
(for $n=0$, we set $\text{W}_{\gamma}[\cdot](x)=1$.).
Then we have, corresponding to \eqref{crumwrons1},
\begin{align}
  \text{W}_{\gamma}[\phi_0,\phi_1,\ldots,\phi_{s-1},\phi_n](x)
  &=\prod_{k=0}^{s-1}\check{\phi}^{[k]}_k(x+i\tfrac{k-s}{2}\gamma)
  \cdot\check{\phi}^{[s]}_n(x),\\
  \check{\phi}^{[s]}_n(x)&\eqdef\frac{\phi^{[s]}_n(x)}
  {\prod_{l=0}^{s-1}\sqrt{V^{[l]}(x+i\tfrac{s-l}{2}\gamma)}}.
\end{align}
Corresponding to \eqref{crumwronsfin}, we obtain
\begin{equation}
  \phi^{[s]}_n(x)=\prod_{l=0}^{s-1}\sqrt{V^{[l]}(x+i\tfrac{s-l}{2}\gamma)}
  \cdot\frac{\text{W}_{\gamma}[\phi_0,\phi_1,\ldots,\phi_{s-1},\phi_n](x)}
  {\text{W}_{\gamma}[\phi_0,\phi_1,\ldots,\phi_{s-1}](x-i\frac{\gamma}{2})},
  \quad(n\geq s\geq 0).
\end{equation}
The proof of the above statements is elementary by induction and the necessary
nontrivial formulas are only
\begin{align}
  &\phi^{[s]}_n(x)=i\frac{\sqrt{V^{[s-1]}(x+i\frac{\gamma}{2})}}
  {\phi^{[s-1]}_{s-1}(x-i\frac{\gamma}{2})}
  \left|\begin{array}{cc}
  \phi^{[s-1]}_{s-1}(x+i\frac{\gamma}{2})&
  \phi^{[s-1]}_n(x+i\frac{\gamma}{2})\\
  \phi^{[s-1]}_{s-1}(x-i\frac{\gamma}{2})&
  \phi^{[s-1]}_n(x-i\frac{\gamma}{2})
  \end{array}\right|,\quad(n\geq s\geq 1),\\[6pt]
  &\left|\begin{array}{cc}
  \text{W}_{\gamma}[f_0,f_1,\ldots,f_{s-1},f_s](x+i\tfrac{\gamma}{2})&
  \text{W}_{\gamma}[f_0,f_1,\ldots,f_{s-1},f_n](x+i\tfrac{\gamma}{2})
  \\[4pt]
  \text{W}_{\gamma}[f_0,f_1,\ldots,f_{s-1},f_s](x-i\tfrac{\gamma}{2})&
  \text{W}_{\gamma}[f_0,f_1,\ldots,f_{s-1},f_n](x-i\tfrac{\gamma}{2})
  \end{array}\right|\n
  =&-i\,\text{W}_{\gamma}[f_0,f_1,\ldots,f_{s-1}](x)\,
  \text{W}_{\gamma}[f_0,f_1,\ldots,f_{s-1},f_s,f_n](x),
  \quad(n\geq s\geq 0).
\end{align}

\section{Summary and Comments}

Since Crum's paper \cite{crum} is crisp and elegant, the underlying
logical structure is not easy to fathom for non-experts or physicists.
In section two we reproduce his results by  a simplest logic in the
language of quantum mechanics and using the factorisation method \cite{infhul},
or the so-called supersymmetric quantum mechanics \cite{susyqm},
so that the similarity and contrast with the corresponding results of
the discrete QM would become clear.
Due to the lack of essential theorems in the theory of difference equations,
some important properties of the spectra and eigenfunctions of the
generic discrete QM must be assumed for the derivation of the associated
Hamiltonian systems in section three.
For example, the hermiticity or the self-adjointness of a discrete QM
Hamiltonian can only be demonstrated after a proper groundstate
wavefunction $\phi_0$ is chosen \cite{os10,os13}.
As mentioned repeatedly, these standard properties are well satisfied
in explicit examples of discrete QM in \cite{os4,os13,os7}.
However, to the best of our knowledge, the general structure of the
solutions of the main difference equation \eqref{maineq} as well as
that for the groundstate \eqref{discphi0def} for generic potential
$V(x)$ and $V^*(x)$, has not yet been investigated in contradistinction
to the ordinary QM.
In this connection, let us mention two interesting examples of
quasi-exactly solvable discrete quantum mechanics (example in \S IIB1
of the first paper of \cite{os10} and in \S3 of \cite{newqes})
in which the oscillation theorem does not hold.

Here are some comments on closely related topics;
shape invariance, orthogonal polynomials and limiting properties to
the ordinary quantum mechanics, etc.

\paragraph{Shape invariance:}
The Hamiltonian may contain several parameters
$\bm{\lambda}=(\lambda_1,\lambda_2,$ $\ldots)$ and we write them explicitly
$\mathcal{H}=\mathcal{H}(\bm{\lambda})$, $V(x)=V(x\,;\bm{\lambda})$,
$\mathcal{A}=\mathcal{A}(\bm{\lambda})$, etc.
Let us consider the case that the potential function of the first
associated Hamiltonian $V^{[1]}(x)=V^{[1]}(x\,;\bm{\lambda})$ has
the same form as the original function $V$ with a different set of
parameters and up to a multiplicative positive constant
$\kappa\in\mathbb{R}_+$ :
\begin{equation}
  V^{[1]}(x\,;\bm{\lambda})=\kappa V(x\,;\bm{\lambda}').
\end{equation}
Here the new set of parameters $\bm{\lambda}'$ is uniquely determined
by $\bm{\lambda}$ (let us write $\bm{\lambda}'=\text{si}(\bm{\lambda})$).
Then this system has the shape invariance \cite{genden,os4,os13},
\begin{equation}
  \mathcal{A}(\bm{\lambda})\mathcal{A}(\bm{\lambda})^{\dagger}
  =\kappa\mathcal{A}(\bm{\lambda'})^{\dagger}
  \mathcal{A}(\bm{\lambda'})
  +\mathcal{E}_1(\bm{\lambda}).
\end{equation}
The shape invariance is a sufficient condition for exact solvability.
The entire energy spectrum and the excited wavefunctions are expressed
in terms of $\mathcal{E}_1(\bm{\lambda})$ and $\phi_0(x\,;\bm{\lambda})$
as follows:
\begin{align}
  &\mathcal{E}_n(\bm{\lambda})=\sum_{s=0}^{n-1}
  \kappa^s\mathcal{E}_1(\bm{\lambda}^{[s]}),\\
  &\phi_n(x\,;\bm{\lambda})\propto
  \mathcal{A}(\bm{\lambda}^{[0]})^{\dagger}
  \mathcal{A}(\bm{\lambda}^{[1]})^{\dagger}
  \mathcal{A}(\bm{\lambda}^{[2]})^{\dagger}
  \cdots
  \mathcal{A}(\bm{\lambda}^{[n-1]})^{\dagger}
  \phi_0(x\,;\bm{\lambda}^{[n]}),
\end{align}
where $\bm{\lambda}^{[n]}$ is $\bm{\lambda}^{[0]}=\bm{\lambda}$,
$\bm{\lambda}^{[n]}=\text{si}(\bm{\lambda}^{[n-1]})$ ($n=1,2,\ldots$).

\paragraph{Orthogonal polynomial:}
Here we consider a generic Hamiltonian \eqref{discrham} of the
discrete QM. That is the shape invariance is not assumed.
Let us define a `real' function $\eta(x)$ ($\eta^*=\eta$) as
a ratio of $\phi_1(x)$ and $\phi_0(x)$,
\begin{equation}
  \frac{\phi_1(x)}{\phi_0(x)}=a+b\,\eta(x),
  \label{etalinear}
\end{equation}
where $a$ and $b$ $(b\neq 0)$ are real constants.
Although $\eta(x)$ is not well defined without specifying $a$ and $b$,
this ambiguity (affine transformation of $\eta(x)$) does not affect
the following discussion.
Then \eqref{V1} implies
\begin{equation}
  V^{[1]}(x+i\tfrac{\gamma}{2})
  =V(x)\,\frac{\eta(x-i\gamma)-\eta(x)}{\eta(x)-\eta(x+i\gamma)}.
  \label{V1Veta}
\end{equation}
Let us assume further that the $n$-th eigenfunction $\phi_n/\phi_0$ is
a degree $n$ polynomial in this $\eta(x)$ for all $n\ge2$:
\begin{equation}
  \phi_n(x)=\phi_0(x)P_n\bigl(\eta(x)\bigr),\quad
  P_n(y)=\sum_{k=0}^na_{n,k}\,y^k,\quad a_{n,k}\in\mathbb{R},
  \quad a_{n,n}\neq 0.
  \label{phin=phi0Pn}
\end{equation}
Obviously $a_{0,0}=1$, $a=a_{1,0}$ and $b=a_{1,1}$.
The orthogonality of the eigenfunctions $\{\phi_n\}$ implies that
$\{P_n\bigl(\eta(x)\bigr)\}$ are orthogonal polynomials in $\eta(x)$
with respect to the weight function $\phi_0(x)^2$.
Then the ratio of the first excited state and the groundstate of
the $s$-th associated Hamiltonian system $\mathcal{H}^{[s]}$ takes
the same form as that of the first associated Hamiltonian system
$\mathcal{H}^{[1]}$ \eqref{etalinear},
\begin{align}
  \frac{\phi^{[s]}_{s+1}(x)}{\phi^{[s]}_{s}(x)}
  =\frac{a_{s+1,s}}{a_{s,s}}+\frac{a_{s+1,s+1}}{a_{s,s}}\,\eta^{[s]}(x),
  \quad
  \eta^{[s]}(x)\eqdef\sum_{k=0}^s\eta(x+i\tfrac{2k-s}{2}\gamma),
\end{align}
and $V^{[s]}$ is related to $V^{[s-1]}$ as in \eqref{V1Veta} and
this goes further down to $V^{[0]}$:
\begin{align}
  V^{[s]}(x+i\tfrac{s}{2}\gamma)
  &=V^{[s-1]}(x+i\tfrac{s-1}{2}\gamma)
  \frac{\eta^{[s-1]}(x+i\tfrac{s-3}{2}\gamma)
  -\eta^{[s-1]}(x+i\tfrac{s-1}{2}\gamma)}
  {\eta^{[s-1]}(x+i\tfrac{s-1}{2}\gamma)
  -\eta^{[s-1]}(x+i\tfrac{s+1}{2}\gamma)}\n
  &=V(x)\prod_{k=0}^{s-1}
  \frac{\eta^{[k]}(x+i\tfrac{k-2}{2}\gamma)
  -\eta^{[k]}(x+i\tfrac{k}{2}\gamma)}
  {\eta^{[k]}(x+i\tfrac{k}{2}\gamma)
  -\eta^{[k]}(x+i\tfrac{k+2}{2}\gamma)}\n
  &=V(x)\prod_{k=0}^{s-1}
  \frac{\eta(x-i\gamma)-\eta(x+ik\gamma)}{\eta(x)-\eta(x+i(k+1)\gamma)}.
\end{align}
In all the known examples of exactly solvable QM (ordinary and discrete),
in which the eigenfunctions have the polynomial form \eqref{phin=phi0Pn},
the function $\eta(x)$ plays a special role to define together with
the Hamiltonian $\mathcal{H}$ an algebraic sufficient condition for exact
solvability. In that case the function $\eta(x)$ is called the
{\em sinusoidal coordinate\/} and the sufficient condition is named
the {\em closure relation\/} \cite{os7,os12,os13}. As shown above,
the shape invariance and the closure relation are very closely related.

\paragraph{Limit from dQM to QM:}
Here we show that the ordinary QM is obtained from the discrete QM
in a certain limit by rescaling parameters.
Let us introduce a positive parameter $c$ and rescale $\gamma$ as
$\gamma/c$ and the parameters in $V(x)$ appropriately,
\begin{align}
  &\mathcal{H}=\mathcal{A}^{\dagger}\mathcal{A}
  =\sqrt{V(x)}\,e^{\frac{\gamma}{c}p}\sqrt{V^*(x)}
  +\!\sqrt{V^*(x)}\,e^{-\frac{\gamma}{c}p}\sqrt{V(x)}-V(x)-V^*(x),\\
  &\mathcal{A}\eqdef i\bigl(e^{\frac{\gamma}{2c}p}\sqrt{V^*(x)}
  -e^{-\frac{\gamma}{2c}p}\sqrt{V(x)}\,\bigr),
  \ \mathcal{A}^{\dagger}\eqdef -i\bigl(\sqrt{V(x)}\,e^{\frac{\gamma}{2c}p}
  -\sqrt{V^*(x)}\,e^{-\frac{\gamma}{2c}p}\bigr).
\end{align}
Assume that $V(x)$ has the following expansion for large $c$,
\begin{equation}
  V(x)=a\Bigl(1+\frac{i\gamma}{c}\,w_1(x)+O\bigl(\frac{1}{c^2}\bigr)\Bigr),
\end{equation}
where $a$ is a positive constant.
Then we have for large $c$:
\begin{align}
  &\frac{c}{\sqrt{a}\,\gamma}\mathcal{A}=\frac{d}{dx}
  -\frac{d\mathcal{W}(x)}{dx}+O\bigl(\frac{1}{c}\bigr),\quad
  \frac{c}{\sqrt{a}\,\gamma}\mathcal{A}^{\dagger}=-\frac{d}{dx}
  -\frac{d\mathcal{W}(x)}{dx}+O\bigl(\frac{1}{c}\bigr),\\
  &\frac{c^2}{a\gamma^2}\,\mathcal{H}
  =p^2+\Bigl(\frac{d\mathcal{W}(x)}{dx}\Bigr)^2
  +\frac{d^2\mathcal{W}(x)}{dx^2}+O\bigl(\frac{1}{c}\bigr),
\end{align}
where the derivative of the pre-potential $\mathcal{W}(x)$ is defined by
$\frac{d\mathcal{W}(x)}{dx}=-\text{Re}\,w_1(x)$.
Therefore ordinary QM is obtained from discrete QM in the $c\to\infty$
limit.

\paragraph{Discrete QM with real shifts:}
Crum's theorem for the discrete QM with real shifts can also be
formulated in a similar manner.
In this case the Hamiltonian is a {\em real symmetric tri-diagonal\/}
(Jacobi) matrix, either of finite or infinite dimensions \cite{os12}.
The factorisation of the positive semi-definite Hamiltonian
$\mathcal{H}=\mathcal{A}^\dagger\mathcal{A}$ also holds and $\mathcal{A}$
consists of the diagonal and super-diagonal elements only and
$\mathcal{A}^\dagger$ being its transpose, which consists of the
diagonal and sub-diagonal elements only.

\paragraph{Connection  with integrable systems:}
Since the Darboux transformation \cite{Darboux} is closely related to 
the inverse scattering method for soliton equations,
it is natural to ask if the present formulation of Crum's theorem
is related to (discrete) integrable systems.
At present, all the explicit examples considered in discrete QM
\cite{os4,os13,os7} (pure imaginary shifts) have an infinite number
of discrete eigenvalues only.
In other words, the corresponding potentials are confining, that is,
they grow to infinity at the boundaries or the spatial infinities.
Thus the free incoming/outgoing waves at infinity do not exist and
the corresponding scattering problem cannot be formulated.
To sum up, we have nothing to report on possible applications of the
present Crum's theorem to (discrete) integrable systems.

\bigskip
After completing this work, we received a recent work by Gaillard
and Matveev \cite{matveev} which has some overlap with the present work.
We thank Vladimir Matveev for sending the new results and for many
useful comments.

\section*{Acknowledgements}

We are supported in part by Grant-in-Aid for Scientific
Research from the Ministry of Education, Culture, Sports, Science and
Technology, No.18340061 and No.19540179.



\begin{thebibliography}{99}
%

\bibitem{crum}
M.\,M.\,Crum,
\JL{Quart. J. Math. Oxford Ser. (2),6,1955,121},
arXiv:physics/9908019.

\bibitem{Darboux}
G. Darboux,
C. R. Acad. Paris \textbf{94} (1882), 1456.

\bibitem{infhul}
L.\,Infeld and T.\,E.\,Hull,
\JL{Rev. Mod. Phys.,23,1951,21}.

\bibitem{susyqm}
See, for example, a review:
F.\,Cooper, A.\,Khare and U.\,Sukhatme,
\PRP{251,1995,267}.

\bibitem{genden}
L.\,E.\,Gendenshtein,
\JL{JETP Lett.,38,1983,356}.

\bibitem{adler}
V.\,\'E.\, Adler,
Theor. Math. Phys. {\bf 101} (1994) 1381.

\bibitem{os4}
S.\,Odake and R.\,Sasaki,
J. Nonlinear Math. Phys. \textbf{12} Suppl. 1 (2005), 507,
arXiv:hep-th/0410102;
\JMP{46,2005,063513},
arXiv:hep-th/0410109.

\bibitem{os12}
S.\,Odake and R.\,Sasaki,
\JMP{49,2008,053503},
arXiv:0712.4106[math.CA].

\bibitem{os13}
S.\,Odake and R.\,Sasaki,
\PTP{119,2008,663},
arXiv:0802.1075[quant-ph].

\bibitem{os6}
S.\,Odake and R.\,Sasaki,
\PTP{114,2005,1245},
arXiv:hep-th/0512155.

\bibitem{os7}
S.\,Odake and R.\,Sasaki,
\JMP{47,2006,102102},
arXiv:quant-ph/0605215;
\PLB{641,2006,112},
arXiv:quant-ph/0605221.

\bibitem{os10}
R.\,Sasaki,
\JMP{48,2007,122104},
arXiv:0708.0702[nlin.SI];
S.\,Odake and R.\,Sasaki,
\JMP{48,2007,122105},
arXiv:0708.0716[nlin.SI].

\bibitem{newqes}
R.\,Sasaki,
J. Nonlinear Math. Phys. \textbf{15} Suppl. 3 (2008), 373,
arXiv:0712.2616[nlin.SI].

\bibitem{askey}
G.\,E.\,Andrews, R.\,Askey and R.\,Roy,
\textit{Special Functions},
Encyclopedia of mathematics and its applications (Cambridge, 1999).

\bibitem{ismail}
M.\,E.\,H.\,Ismail
\textit{Classical and quantum orthogonal polynomials in one variable},
Encyclopedia of mathematics and its applications (Cambridge, 2005).

\bibitem{koeswart}
R.\,Koekoek and R.\,F.\,Swarttouw,
arXiv:math.CA/9602214.

\bibitem{qoscill}
A.\,J.\, Macfarlane,
\JP{A22,1989,4581};
L.\,C.\, Biedenharn,
\JP{A22,1989,L873};
C.-P.\, Sun and H.-C.\, Fu,
\JP{A22,1989,L983}.

\bibitem{os11}
S.\, Odake and R.\, Sasaki,
\PLB{663,2008,141},
arXiv:0710.2209[hep-th].
 
\bibitem{kos}
R.\, Kobayashi, S.\, Odake and R.\, Sasaki,
in preparation.

\bibitem{birth}
R.\,Sasaki,
J. Math. Phys. October issue (2009) in press.

\bibitem{os16}
S.\,Odake and R.\,Sasaki,
\PLB{679,2009,414},
arXiv:0906.0142[math-ph].

\bibitem{os17}
S.\,Odake and R.\,Sasaki,
Preprint DPSU-09-4, YITP-09-52.

\bibitem{matveev}
P. Gaillard and V.\,B.\, Matveev,
``Wronskian and Casorati determinant representation for
Darboux-P\"oschel-Teller potentials and their difference extensions,"
Preprint RIMS-1653, January 2009.

\end{thebibliography}
\end{document}